\documentclass[preprint,12pt,authoryear]{elsarticle}
\usepackage{amsmath}
\usepackage{graphicx}
\usepackage{booktabs}
\usepackage[table]{xcolor}
\usepackage{listings}
\usepackage{enumerate}
\usepackage{natbib}
\usepackage[a4paper]{geometry}
\usepackage{graphicx}
\usepackage[textwidth=8em,textsize=small]{todonotes}
\usepackage{amsmath}
\usepackage{natbib}
\usepackage{array}
\usepackage[utf8]{inputenc}
\usepackage{graphicx}
\usepackage{longtable}
\graphicspath{ {./images/} }
\usepackage{multirow}
\usepackage{float}
 \usepackage{xspace}
 \usepackage{appendix}
 \usepackage{array}
\usepackage{amssymb}
\usepackage{dirtytalk}
\usepackage{etoolbox}
\usepackage{url}
\newcolumntype{C}[1]{>{\centering\arraybackslash}p{#1}}

\usepackage{booktabs}

\usepackage{url} 
\date{}
\journal{ARXIV}

\begin{document}

\begin{frontmatter}

\title{Mortality Modelling using
Generalized Estimating Equations   
}
\author{Reza Dastranj\corref{cor1}}

\ead{dastranj@math.muni.cz}

\author{Martin Kol\'a\v r}
\ead{mkolar@math.muni.cz}

\address{Department of Mathematics and Statistics, Masaryk University, Kotlářská 2, 611 37 Brno, Czech Republic}

\cortext[cor1]{Corresponding Author}

\begin{abstract}

This paper presents an application of Generalized Estimating Equations (GEE) for analyzing age-specific death rates (ASDRs), constituting a longitudinal dataset with repeated measurements over time. GEE models, known for their robustness in handling correlated data, offer a reliable solution when individual data records lack independence, thus violating the commonly assumed independence and identically distributed (iid) condition in traditional models. In the context of ASDRs, where correlations emerge among observations within age groups, two distinct GEE models for single and multipopulation ASDRs are introduced, providing robust estimates for regression parameters and their variances. We explore correlation structures, encompassing independence, AR(1), unstructured, and exchangeable structures, offering a comprehensive evaluation of GEE model efficiency in both single and multipopulation ASDRs. We critically examines the strengths and limitations of GEE models, shedding light on their applicability for mortality forecasting. Through detailed model specifications and empirical illustrations, the study contributes to an enhanced understanding of the nuanced capabilities of GEE models in predicting mortality rates.

\end{abstract}

\begin{keyword}
Mortality forecasting \sep Quasi-likelihood\sep Generalized estimating equations \sep
Longitudinal analysis \sep Random walks with drift.



\end{keyword}

\end{frontmatter}

\section{Introduction}\label{sec:intro}
ASDRs serve as vital indicators of mortality trends, collected sequentially over different years for various populations. Within the same age group, mortality rates exhibit higher similarity, forming a longitudinal dataset with inherent correlations \citep{dastranj2023age}.

Traditional statistical approaches, such as generalized linear models (GLMs), assume independence among individual rows in the data \citep{nelder1972generalized, mccullagh2019generalized,james2013introduction}. However, when dealing with longitudinal and clustered data, especially in mortality rates within the same age group, this assumption falls short, violating the iid requirement. To address these limitations, statisticians developed methods, leading to the emergence of GEE as a powerful tool explicitly designed to extend the GLM algorithm. GEE provides a robust framework for modelling correlated data, particularly in scenarios where straightforward GLM methods may fall short \citep{liang1986longitudinal, hardin2002generalized,lee2004conditional}.

This paper delves into the powers and properties of GEE in the context of mortality forecasting. Two distinct models are presented: one tailored for multipopulation scenarios and another designed for single-population studies. The multipopulation model incorporates key predictors such as country, gender, age, and their intricate interactions with mortality covariates $k_t$, $k_t^2$, and cohort.

In the landscape of statistical modelling, Linear Mixed Effects Models (LMEs) and GEEs represent two prominent approaches for handling correlated data.

Subject-Specific Models (LMEs): LMEs focus on estimating both fixed effects and variance components, providing insights into the magnitude of each source of variation in the data \citep{pinheiro2006mixed}. They are particularly suitable when there is interest in understanding the underlying population and capturing the variability attributed to different grouping factors. However, subject-specific models have a conditional formulation with random intercepts and slopes, making it challenging to provide a population-average interpretation of the model parameters. Estimates obtained from subject-specific models are specific to the individuals or subjects under consideration \citep{lee2004conditional}.

In the LME model \citep{dastranj2023age} , designating "country" as a random effect signifies a specific interest in the underlying population of the six European countries included in the dataset. The variance of the random effect for the country quantifies the variability in the response variable attributed to differences between these observed countries. When exploring the variance of the random effect for countries beyond the dataset, the analysis considers how much the response variable may vary for countries not explicitly observed. This approach facilitates a broader understanding of the potential variability in mortality rates across a larger population, extending insights beyond the specific countries included in the dataset.

Marginal Models (GEEs): In contrast, GEEs prioritize estimating regression coefficients without explicit interest in variance components. They are robust models designed to handle correlated data and are especially useful when the primary concern is modelling the mean response while treating within-group correlation as a nuisance. Marginal models aim to capture the population-average effects \citep{lee2004conditional}, allowing for a more generalized interpretation of the model parameters. However, they may encounter challenges when dealing with ordinal data, particularly in specifying an underlying distribution. This is especially evident when GEEs are applied, as they focus on estimating population-average effects while accommodating correlated data.

\section{Modelling Mortality Rates in Single and Multi-Populations}\label{sec2}

\begin{figure}[H]
    \centering
    \includegraphics[width=\textwidth]{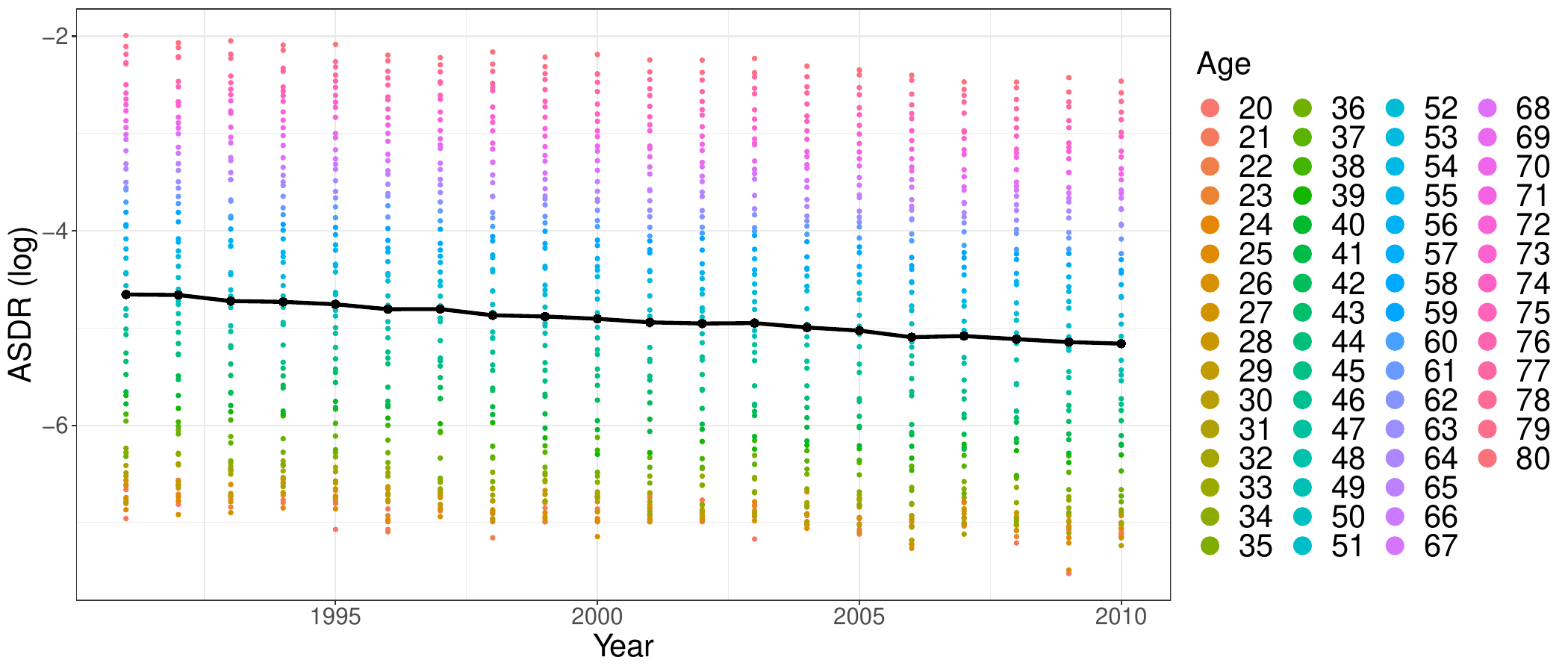}
    \caption{Plot of $k_t$: Average of log ASDRs across all age groups of Czech males ($1991-2010$).}
    \label{fig1}
\end{figure}

In Figure \ref{fig1}, the thick black curve represents $k_t$, obtained by taking the average of log mortality rates across ages. For example, $k_{1991}$ is the 1991 average of log ASDRs. $k_t$ is the overall age-specific mortality pattern (trajectory) of the Czech male dataset\citep{HMD}. 
\subsection{The Model Formulation}\label{sec2.1}

The foundational model structure for GEEs in the context of mortality rates and age groups can be expressed as:

\begin{equation}
  y_{xt} = \beta_0 + \sum_{k} X_{xtk}\beta_k + \text{CORR} + \text{error}.  
\end{equation}

Within this formulation, $\beta$'s represent fixed effects, while "CORR" encapsulates the correlation structure inherent in GEE. For mortality rates within specific age groups, we assume we have $y_{xt}$ (log-transformed mortality rates) as our outcome for individual age group $x$ at time $t$. This could just as easily represent individuals within other types of levels, like genders or counties. Additionally, we have $X_{xt}$, the matrix of predictors. 

In the context of repeated observations within age groups over time, it is crucial to address the potential biases in standard error estimations for both time-varying and time-invariant predictors. Specifically, conventional models not incorporating the correlation within age groups over time tend to exhibit an overestimation of standard errors for $\beta$'s associated with time-varying predictors. Simultaneously, there is a tendency for the standard errors of time-invariant predictors to be underestimated. This phenomenon underscores the necessity of employing advanced modelling techniques, such as GEE, which explicitly account for temporal dependencies and yield more accurate standard error estimates. GEE's capacity to appropriately handle the correlation structure within longitudinal data, especially within age groups, enhances the reliability of parameter estimates, contributing to robust statistical inference in the presence of repeated observations.

\subsection{Correlation Structures}

GEEs require specifying a correlation structure for repeated measures. Several commonly used structures include:

\paragraph{Independence:}
The Independence structure assumes no correlation between repeated measures, similar to Ordinary Least Squares (OLS). The corresponding correlation matrix is represented as:
\[
\begin{bmatrix}
1 & 0 & \cdots &0 \\
0 & 1 & \cdots & 0 \\
\vdots &\vdots &\ddots &\vdots\\
0 & 0 &  \cdots &1 \\
\end{bmatrix}
\]

The method of independence estimating equations incorrectly assumes that observations within a subject are independent, treating correlated responses as if they were independent \citep{pan2002selecting}. When specifying an independence correlation structure for repeated measures in GEEs, it essentially assumes zero correlation between repeated measures. This implies independence, rendering the GEE model behavior akin to a GLM. The specification of an independence correlation structure treats each observation as entirely independent of others, resembling a standard GLM. It neglects potential correlations or dependencies between observations, resulting in a simpler and more straightforward model concerning the correlation structure.

\paragraph{AR(1):}
The Autoregressive (AR(1)) structure assumes decreasing correlation with increasing time separation. The corresponding correlation matrix is:
\[
\begin{bmatrix}
1 & \rho & \cdots&\rho^{n-1} \\
\rho & 1 & \cdots &\rho^{n-2} \\
\vdots &\vdots &\ddots &\vdots\\
\rho^{n-1} & \rho^{n-2} & \cdots&1 \\
\end{bmatrix}
\]

\paragraph{Exchangeable:}
Exchangeable, or compound symmetry, assumes all correlations between repeated measures are the same. The correlation matrix for this structure is given by:
\[
\begin{bmatrix}
1 & \rho & \cdots&\rho \\
\rho & 1 & \cdots&\rho \\
\vdots &\vdots &\ddots &\vdots\\
\rho & \rho & \cdots&1 \\
\end{bmatrix}
\]

\paragraph{Unstructured:}
The Unstructured structure is the most flexible, allowing different correlations between all pairs of repeated measures. The correlation matrix is represented as:
\[
\begin{bmatrix}
1 & \rho_{12} & \cdots& \rho_{1n} \\
\rho_{21} & 1&\cdots & \rho_{2n} \\
\vdots &\vdots &\ddots &\vdots\\
\rho_{n1} & \rho_{n2} &\cdots& 1 \\
\end{bmatrix}
\]

Each $\rho$ represents a correlation coefficient, capturing the strength and direction of the correlation between respective measures. These structures provide flexibility in modelling the dependence between repeated measures, catering to different assumptions about the nature of the correlation in longitudinal data.

In comparing the fit of GEE models with different correlation structures, the Quasi-Likelihood Information Criterion (QIC) serves as a robust measure. QIC balances goodness of fit and model complexity, guiding the selection of the most suitable model. Unlike traditional criteria like Akaike Information Criterion (AIC) and Bayesian Information Criterion (BIC), which may not align well with GEE's quasi-likelihood framework, QIC is specifically tailored for accurate model comparison in this context. Its variants, like QICu, offer additional corrections for overdispersion. Therefore, when dealing with GEE models for mortality data, QIC stands out as a more appropriate and reliable criterion, ensuring precise model selection.

The method of GEE models is now well established. Let $m_{c,g,x,t}$ denote the ASDR at age $x$ and time $t$ of gender $g$ in country $c$, for $c=$ $1, 2,\cdots,M$; $g=1, 2$; $x = 0, 1, \cdots , \omega$; and $t=0,1, \cdots, T$ \citep{dickson2019actuarial, macdonald2018modelling}. Let $y_{cgxt}= \log (m_{c,g,x,t})$. Let $k_t$ denote the average of $y_{cgxt}$ at time $t$ across all age groups in all countries and genders:
\begin{align}
k_{t} = \dfrac{\sum\limits_{c=1}^{M}\sum\limits_{g=1}^{2}\sum\limits_{x=0}^{\omega} y_{cgxt}}{2 M(\omega+1)},
\label{ali6}
\end{align}
 for $t=0, 1,  \cdots, T$. $k_{t}$ will be included as predictor in our model. We will employ random walks with drift to forecast the future values of $k_t$ \citep{dastranj2023age}. We refer to this covariate as the \say{mortality covariate} due to its significance in capturing the temporal patterns of mortality rates. We propose the development of a GEE model to capture the relationships between the log ASDRs and the mortality covariate. Within this framework, the mortality covariate is considered as the driver or exogenous series, while the log ASDRs is treated as dependent variable influenced by this driver.

The GEE models, implemented using the geeglm function from the geepack package in R \citep{hojsgaard2006r}, for both single and multi-population scenarios, are defined as follows:

For single population:

\begin{verbatim}
geeglm(y ~ age + age:I(kt) + age:I(kt^2) + age:cohort,
           id = age, waves = year, corstr = "exchangeable",
           weights = agenum/mean(agenum), data = ASDRs)
\end{verbatim}

For multi population:

\begin{verbatim}
geeglm(y ~ country + gender + age + 
           age:I(kt) + age:I(kt^2) + age:cohort,
           id = country:gender:age, waves = year, 
           corstr = "exchangeable", weights = agenum/mean(agenum), 
           data = ASDRs)

\end{verbatim}

These models incorporate various predictors, including age, country, gender, and their interactions, to capture the complexity of mortality rates within the specified populations. The selection of the correlation structure (corstr) is pivotal in accounting for temporal dependencies within the data, specifically recognizing the potential correlation of mortality rates over consecutive years. By choosing an appropriate correlation structure, such as autoregressive ("ar1"), the models acknowledge and adjust for these temporal dependencies. This consideration significantly enhances the reliability of parameter estimates, contributing to robust and accurate mortality forecasting in both single and multi-population scenarios. The corstr argument in the geeglm function is a character string that designates the correlation structure for the GEE model. The permissible options include: "independence",
"exchangeable",
"ar1","unstructured", and "userdefined" (allows the specification of a user-defined correlation structure).

These models can be mathematically represented as:

For the single population:
\begin{equation}\label{eq3}
    y_{xt} = a_x + b_xk_t + c_xk_t^2 + \gamma_x(t-x) + \epsilon_{xt}
\end{equation}
Here, $a_x$ denotes the age effect of the intercept, while $b_x$ and $c_x$ represent the age effects of $k_t$ and $k_t^2$, respectively. The term $\gamma_x$ denotes the cohort effect, and $\epsilon_{xt}$ represents the error term. The term $\gamma_{x}$ demonstrates a modulating pattern contingent on age, reflecting the influence of birth cohorts on mortality rates. The cohort effect suggests that individuals born in distinct years (cohorts) may experience varying mortality rates due to unique historical, societal, or environmental factors relevant to their birth years. Our conceptualization draws inspiration from \citet{renshaw2006cohort}, who introduced a cohort factor in the Lee-Carter \citep{lee1992modeling} model, coupled with an age-modulating coefficient (see, \citealt{booth2008mortality}). In our GEE model, we explicitly integrate this cohort effect, configuring the model to comprehensively consider age-related influences.

For the multi-population scenario:
\begin{equation}
    y_{cgxt} = a_c + a_g + a_x + b_{cgx}k_t + c_{cgx}k_t^2 + \gamma_{cgx}(t-x) + \epsilon_{cgxt}
\end{equation}
In this formulation, $a_c$, $a_g$, and $a_x$ denote the country, gender, and age effects in the intercept, respectively. Additionally, $b_{cgx}$ and $c_{cgx}$ represent the country-gender-age effects of $k_t$ and $k_t^2$, and $\gamma_{cgx}$ denotes the cohort effect. The error term is denoted by $\epsilon_{cgxt}$.

\cite{dastranj2023age} have demonstrated the rationale behind incorporating the quadratic term ($k_t^2$) in the model when utilizing the mortality covariate $k_t$. Our analysis revealed that the relationship between $y_{xt}$ and $k_t$ exhibits a quadratic trend, suggesting a more accurate representation of the underlying dynamics. The plotted graph of $y_{xt}$ against $k_t$ revealed a combination of a quadratic trend and stochastic variation, reinforcing the justification for introducing the quadratic term in the model.

\subsection{Single Population Analysis - Czech Republic (Males Age 20-80)}

Focusing on Czech males aged 20 to 80 during 1991-2010, we employ GEE model \ref{eq3} with independence (geeInd), exchangeable (geeEx), AR(1) (geeAr1), and unstructured (geeUns) correlation structures. Table \ref{tab1} shows QIC values for model selection. Despite close QICs, geeInd and geeEx feature lower Corrected Information Criterion (CIC) values. We choose geeEx as the final model for predicting mortality rates in Czech males from 2011 to 2019 (pre-COVID-19).

\begin{table}[H]
  \centering
  \caption{QIC Values for Models with Different Correlation Structures}
  \label{tab:qic-values}
  \begin{tabular}{lcccccc}
    \toprule
    \textbf{Model} & \textbf{QIC} & \textbf{QICu} & \textbf{Quasi Lik} & \textbf{CIC}  & \textbf{QICC} \\
    \midrule
    geeInd & $5.01 $ & $493$ & $-0.251$ & $2.27 \times 10^{-23}$  & $-645$ \\
    geeEx & $5.01 $ & $493$ & $-0.251$ & $2.46 \times 10^{-23}$  & $-647$ \\
    geeAr1 & $5.01 $ & $493$ & $-0.251$ & $2.01 \times 10^{-14}$ & $-647$ \\
    geeUns & $5.09 $ & $493$ & $-0.255$ & $2.87 \times 10^{-5}$  & $-1000$ \\
    \bottomrule
  \end{tabular}
  \label{tab1}
\end{table}

\subsection{Multi-Population Analysis - Austria and Czech Republic (Females and Males, Age 20-80)}

This section explores mortality rates for Austria and Czech Republic females and males aged 20 to 80 during 1991-2010, serving as the training set. Utilizing three GEE models (geeInd, geeEx, geeAr1) with independence, exchangeable, and AR(1) correlation structures, Table \ref{tab2} displays QIC values for model selection. Opting for geeEx, incorporating an exchangeable structure, this final model predicts mortality rates for the four populations during 2011-2019. While QIC and CIC values for Independence and Exchangeable structures are comparable, the exchangeable structure is deemed more suitable for ASDRs due to its assumption that the correlation between data points remains consistent over time, aligning with the nature of ASDR trends.

\begin{table}[H]
  \centering
  \caption{QIC Values for Models with Different Correlation Structures (Multipopulation).}
  \label{tab:qic-values}
  \begin{tabular}{lcccccc}
    \toprule
    \textbf{Model} & \textbf{QIC} & \textbf{QICu} & \textbf{Quasi Lik} & \textbf{CIC} & \textbf{Params} & \textbf{QICC} \\
    \midrule
    geeInd & 372.6 & 630.4 & -69.2 & 117.1 & 246 & -30255.4 \\
    geeEx & 372.6 & 630.4 & -69.2 & 117.1 & 246 & -30255.4 \\
    geeAr1 & 623.5 & 635.5 & -71.8 & 240.0 & 246 & -30004.5 \\
    \bottomrule
  \end{tabular}
   \label{tab2}
\end{table}

\begin{figure}[H]
    \centering
    \includegraphics[width=\textwidth]{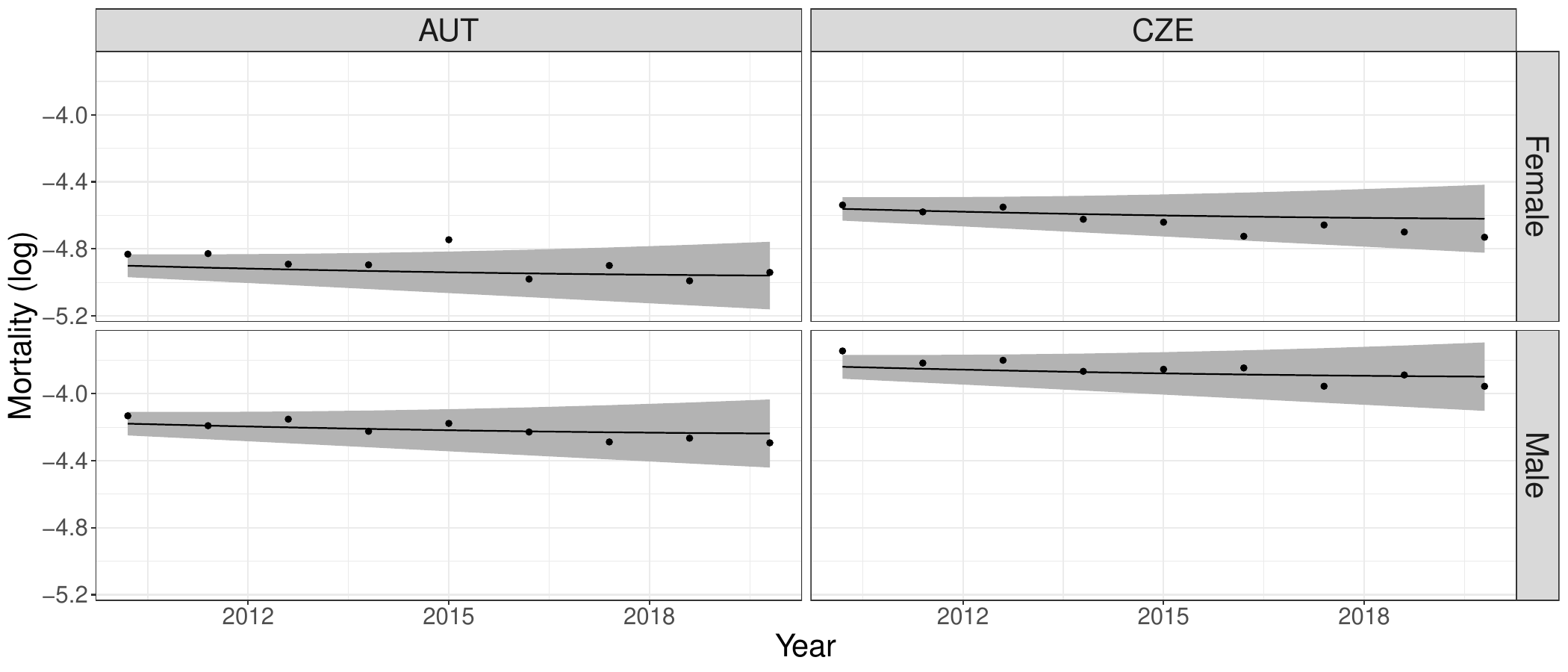}
    \caption{Predictions of Multipopulation GEE Model for ASDRs in Austria and Czech Republic: Age Group 65 (2011-2019).}
    \label{fig2}
\end{figure}

Figure \ref{fig2} illustrates the predictions of the multipopulation GEE model applied to ASDRs in Austria and the Czech Republic, categorized by gender and age group. The analysis focuses on the age group 65 over the years 2011 to 2019, with a training set encompassing the years 1999 to 2010. The GEE model is employed to account for the longitudinal nature of the dataset, addressing correlated observations inherent in longitudinal studies.

The gray-shaded bands in the plot represent $95\%$ prediction intervals, calculated using standard errors derived from the Quasi-likelihood method. It's essential to note that these standard errors differ from those obtained through likelihood-based methods, such as those employed in GLM. The application of GEE is particularly pertinent for longitudinal datasets, as it enables effective modelling of correlated data, acknowledging the temporal dependencies inherent in ASDR observations.

This approach facilitates a nuanced understanding of mortality trends, offering a robust framework for capturing the intricate relationships within the dataset. The exchangeable correlation structure, coupled with the prediction intervals derived from Quasi-likelihood standard errors, enhances the reliability of the predictions and provides valuable insights into the mortality patterns of the specified populations.

The R code for fitting and forecasting the GEE models to mortality data is available in a GitHub repository maintained by the first author of this paper. Specifically, the code covers the case of a single population for males in the Czech Republic and the case of multi-populations for both males and females in Austria and the Czech Republic.

\section{Conclusion}

This study explores the utility of GEEs as an efficient tool for modelling mortality rates. By conducting a thorough analysis of longitudinal mortality datasets, GEEs demonstrate their computational simplicity, particularly in effectively handling categorical data and correlated observations. However, we need to be careful about the inherent limitations in GEE modelling. The absence of a fully specified likelihood function and constraints in likelihood-based methods may impact the extent of statistical analysis achievable with GEEs. Additionally, cautious interpretation is warranted, especially concerning empirical-based standard errors, which may lead to underestimation, particularly in smaller sample sizes. Striking a balance between computational efficiency and nuanced model specifications becomes pivotal when utilizing GEEs in mortality rate modelling. 

Despite their limitations, GEEs hold the promise of advancing our understanding of mortality trends and enhancing the quality of studies in this field.

\bibliographystyle{chicago}
\bibliography{mybib}

\begin{thebibliography}{}

\bibitem[\protect\citeauthoryear{Booth and Tickle}{Booth and Tickle}{2008}]{booth2008mortality}
Booth, H. and L.~Tickle (2008).
\newblock Mortality modelling and forecasting: A review of methods.
\newblock {\em Annals of actuarial science\/}~{\em 3\/}(1-2), 3--43.

\bibitem[\protect\citeauthoryear{Dastranj and Kolar}{Dastranj and Kolar}{2023}]{dastranj2023age}
Dastranj, R. and M.~Kolar (2023).
\newblock Age-gender-country-specific death rates modelling and forecasting: A linear mixed-effects model.
\newblock {\em arXiv preprint arXiv:2311.18668\/}.

\bibitem[\protect\citeauthoryear{Dickson, Hardy, and Waters}{Dickson et~al.}{2019}]{dickson2019actuarial}
Dickson, D.~C., M.~R. Hardy, and H.~R. Waters (2019).
\newblock {\em Actuarial mathematics for life contingent risks}.
\newblock Cambridge University Press.

\bibitem[\protect\citeauthoryear{Hardin and Hilbe}{Hardin and Hilbe}{2002}]{hardin2002generalized}
Hardin, J.~W. and J.~M. Hilbe (2002).
\newblock {\em Generalized estimating equations}.
\newblock chapman and hall/CRC.

\bibitem[\protect\citeauthoryear{HMD}{HMD}{2022}]{HMD}
HMD (2022).
\newblock {Human Mortality Database. Max Planck Institute for Demographic Research (Germany), University of California, Berkeley (USA), and French Institute for Demographic Studies (France)}.
\newblock Available at \url{www.mortality.org} or \url{www.humanmortality.de} (data downloaded on 09-05-2022].

\bibitem[\protect\citeauthoryear{H{\o}jsgaard, Halekoh, and Yan}{H{\o}jsgaard et~al.}{2006}]{hojsgaard2006r}
H{\o}jsgaard, S., U.~Halekoh, and J.~Yan (2006).
\newblock The r package geepack for generalized estimating equations.
\newblock {\em Journal of statistical software\/}~{\em 15}, 1--11.

\bibitem[\protect\citeauthoryear{James, Witten, Hastie, Tibshirani, et~al.}{James et~al.}{2013}]{james2013introduction}
James, G., D.~Witten, T.~Hastie, R.~Tibshirani, et~al. (2013).
\newblock {\em An introduction to statistical learning}, Volume 112.
\newblock Springer.

\bibitem[\protect\citeauthoryear{Lee and Carter}{Lee and Carter}{1992}]{lee1992modeling}
Lee, R.~D. and L.~R. Carter (1992).
\newblock Modeling and forecasting {U.S.} mortality.
\newblock {\em Journal of the American statistical association\/}~{\em 87\/}(419), 659--671.

\bibitem[\protect\citeauthoryear{Lee and Nelder}{Lee and Nelder}{2004}]{lee2004conditional}
Lee, Y. and J.~A. Nelder (2004).
\newblock Conditional and marginal models: another view.

\bibitem[\protect\citeauthoryear{Liang and Zeger}{Liang and Zeger}{1986}]{liang1986longitudinal}
Liang, K.-Y. and S.~L. Zeger (1986).
\newblock Longitudinal data analysis using generalized linear models.
\newblock {\em Biometrika\/}~{\em 73\/}(1), 13--22.

\bibitem[\protect\citeauthoryear{Macdonald, Richards, and Currie}{Macdonald et~al.}{2018}]{macdonald2018modelling}
Macdonald, A.~S., S.~J. Richards, and I.~D. Currie (2018).
\newblock {\em Modelling mortality with actuarial applications}.
\newblock Cambridge University Press.

\bibitem[\protect\citeauthoryear{McCullagh and Nelder}{McCullagh and Nelder}{1989}]{mccullagh2019generalized}
McCullagh, P. and J.~Nelder (1989).
\newblock {\em Generalized Linear Models. 2nd Edition}.
\newblock Chapman and Hall, London.

\bibitem[\protect\citeauthoryear{Nelder and Wedderburn}{Nelder and Wedderburn}{1972}]{nelder1972generalized}
Nelder, J.~A. and R.~W. Wedderburn (1972).
\newblock Generalized linear models.
\newblock {\em Journal of the Royal Statistical Society Series A: Statistics in Society\/}~{\em 135\/}(3), 370--384.

\bibitem[\protect\citeauthoryear{Pan and Connett}{Pan and Connett}{2002}]{pan2002selecting}
Pan, W. and J.~E. Connett (2002).
\newblock Selecting the working correlation structure in generalized estimating equations with application to the lung health study.
\newblock {\em Statistica Sinica\/}, 475--490.

\bibitem[\protect\citeauthoryear{Pinheiro and Bates}{Pinheiro and Bates}{2006}]{pinheiro2006mixed}
Pinheiro, J. and D.~Bates (2006).
\newblock {\em Mixed-Effects models in S and S-PLUS}.
\newblock Springer Science \& Business Media.

\bibitem[\protect\citeauthoryear{Renshaw and Haberman}{Renshaw and Haberman}{2006}]{renshaw2006cohort}
Renshaw, A.~E. and S.~Haberman (2006).
\newblock A cohort-based extension to the {Lee--Carter} model for mortality reduction factors.
\newblock {\em Insurance: Mathematics and Economics\/}~{\em 38\/}(3), 556--570.

\end{thebibliography}

\end{document}